\documentclass[useAMS,usenatbib]{mn2e}
\usepackage{graphicx}
\usepackage{txfonts}
\begin{document}

\title[Spectrum evolution in PSR B1259$-$63
and gigahertz-peaked spectra.]{Spectrum evolution in binary pulsar B1259$-$63/LS 2883 Be~star
and gigahertz-peaked spectra.}
\author[J. Kijak et al.]{J. Kijak$^{1}$\thanks{E-mail: jkijak@astro.ia.uz.zgora.pl},
M. Dembska$^{1}$, W. Lewandowski$^{1}$, G. Melikidze$^{1,2}$ and M. Sendyk$^{1}$ \\
$^1$ Kepler Institute of Astronomy, University of Zielona G\'ora , Lubuska
2, 65-265 Zielona G\'ora, Poland\\
$^2$ Ilia State University, E. Kharadze Abastumani
Astrophysical Observatory, Tbilisi, Georgia}
\date{Accepted . Received ; in original form }
\maketitle

\begin{abstract}

We study the radio spectrum of PSR B1259$-$63 orbiting around the Be star LS~2883 and show that the shape of the
spectrum depends on the orbital phase. At frequencies below 3~GHz PSR B1259$-$63 flux densities are lower when measured
near the periastron passage than those measured far from periastron. We suggest that an interaction of the radio waves
with the Be star environment accounts for this effect. While it is quite natural to explain the pulsar eclipse by the
presence of an equatorial disk around LS~2883, this disk alone cannot be responsible for the observed spectral
evolution of PSR B1259$-$63 and we, therefore, propose a qualitative model which explains this evolution. We consider
two mechanisms that might influence the observed radio emission: free-free absorption and cyclotron resonance. We
believe that this binary system can hold the clue to the understanding of gigahertz-peaked spectra of pulsars.

\end{abstract}

\label{firstpage}

\begin{keywords}
{(stars:) pulsars: general -- radiation mechanism: non-thermal -- pulsars: individual: B1259$-$63,
B1054$-$62, J1809$-$1917,B1822$-$22, B1823$-$13, B1828$-$11, -- stars: winds, outflows}

\end{keywords}

\section{Introduction}
Generally, the observed radio spectra of most pulsars can be modelled as a power law with negative spectral indices of about
$-1.8$ \citep{maron00}. If a pulsar can be observed at frequencies low enough (i.e. 100 - 600 MHz) it may also show a low
frequency turn-over in its spectrum \citep{sieb73,mal94}.

On the other hand \citet{lori95} mentioned three pulsars which have positive spectral indices in the frequency range $300 -
1600$ MHz. Later \citet{maron00} reexamined spectra of these pulsars taking into account the data obtained at higher
frequencies (above $1.6$ GHz) and consequently were the first to demonstrate a possible existence of spectra with turn-over at
high frequencies, about 1 GHz. Motivated by this unusual spectral feature \citet{km04} selected several pulsars which showed a
decreasing flux density at frequencies below 1 GHz. \citet{kijak07} used multifrequency flux measurements for these candidate
pulsars and presented the first direct evidence of a high frequency turn-over. A frequency at which such a spectrum shows the
maximum flux was called the peak frequency.

Based on their observations of these pulsars \citet{kijak11} provided a definite evidence for a new type of pulsar radio
spectra. These spectra show the maximum flux above 1 GHz, while at higher frequencies the spectra look like a typical pulsar
spectrum. At lower frequencies (below 1 GHz) the observed flux decreases, showing a positive spectral index. They called these
objects the gigahertz-peaked spectra (GPS) pulsars.

 \citet{kijak11} also indicated that the GPS pulsars are relatively young
objects and they usually adjoin such interesting environments as HII regions or compact pulsar wind nebulae (PWN).
Additionally they seem to be coincident with the known but sometimes unidentified X-ray sources from 3rd EGRET
Catalogue or HESS observations. We can assume that the GPS appearance owes to the environmental conditions around the
neutron stars rather than to the radio emission mechanism. The issue of our interest PSR B1259$-$63 was also inscribed
by \citet{lori95} in the list of pulsars with positive spectral indices. Therefore, B1259$-$63 seems a natural
candidate to be classified as the GPS pulsar. This pulsar was discovered by \citet{joh92} in 1990 during a large-scale
high frequency ($1.5$~GHz) survey of the Galactic plane. It is important to mention that PSR B1259$-$63 is the only
known radio pulsar orbiting a massive main-sequence Be star.

PSR B1259$-$63 has a short period of 48 ms and a characteristic age of 330 kyr. Its average DM is about 147 pc cm$^{-3}$ and
the corresponding distance is about 2.75 kpc. The companion star LS~2883 is a 10-mag massive Be star with a mass of about 10
M$_{\sun}$ and a radius of 6 R$_{\sun}$. Let us note that Be stars are generally believed to have a hot tenuous polar wind and
a cooler high density equatorial disk. \citet{joh94} and \citet{mjm95} suggested the presence of the disk around LS~2883. Its
density is quite high near the star and falls off as a power law with the distance from the star. The disk is likely to be
highly tilted with respect to the pulsar orbital plane, and PSR B1259$-$63 is eclipsed for about 35 days as it goes behind the
disk. The pulsar has a long orbital period of 1237 days and a large eccentricity of 0.87 with a projected semi-major axis,
$a\sin i$ of 2.6 AU.

The PSR B1259$-$63/LS 2883 is a unique binary system, which emits unpulsed non-thermal emission over a wide range
of frequencies ranging between radio and $\gamma-$rays and its flux varies with orbital phase. {\it Chandra} observed
this system in 2009 (shortly after the apastron passage) and failed to detect any pulsation with a period equal to that
of the pulsar \citep{pav11}. The broadband emission from this system is generally believed to be produced by the
ultra-relativistic electrons accelerated in the bow-shock that appears while the relativistic pulsar wind (PW) collides
with the dense non-relativistic outflow of the companion star, producing the Pulsar Wind Nebula (PWN) \citep{kong11}.

In this paper we analyse the available published data of PSR B1259$-$63 radio flux measured during three periastron passages
(in 1997, 2001 and 2004) and present detailed study of the pulsar spectrum.

\section{Pulsars with the GPS}

To compare various shapes of the B1259$-$63 spectrum observed at different orbital phases with that of the GPS pulsars
we use the same fitting method.
This method was previously used by \citet{kuzm01} for pulsar spectra with turn-over at low frequencies. Using the data
of flux measurements with errors for several GPS pulsars (or at least showing a signature of this unusual spectral
feature) taken from \citet{kijak07} and \citet{kijak11} we analysed the spectra using a function
\begin{equation}
F(\nu)=10^{(ax^2+bx+c)},\ \ \ {\rm here}\  x \equiv \log_{10}{\nu}.
\end{equation}
We carried out fitting of the data using an implementation of
the nonlinear least-squares Marquardt-Levenberg algorithm.

\begin{figure}
\includegraphics[width=8cm]{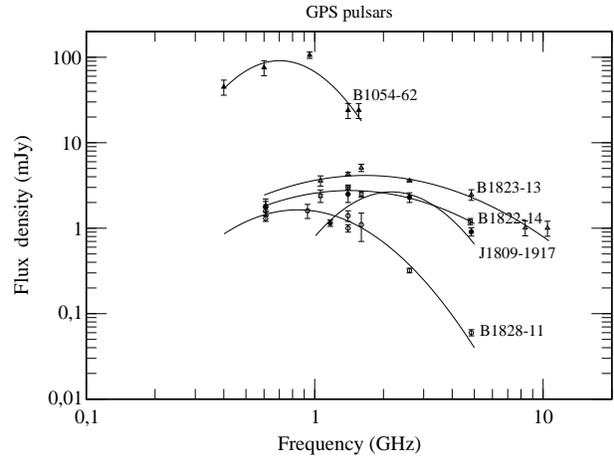}
\caption{The spectra of five known GPS pulsars. The flux measurements are taken from \citet{kijak07, kijak11},
and the curves represent our fits to the data points (see Table~1).}
\label{fig1}
\end{figure}

In Table 1 we present the values of fitting parameters $a$, $b$ and $c$,  the reduced $\chi^2$, the peak frequency
calculated using the method described above and $\nu_{p}^{K}$ are taken from \citet{kijak11}. The fitted spectra are
presented in Fig.~\ref{fig1}.
\begin{table}
\caption{The results of our spectral fits to the known GPS pulsars. The values of peak frequencies $\nu_{p}^{K}$ are
taken from \citet{kijak11}. $\nu_{p}$ is estimated using different spectral fit methods (see text for details). 
In the first column (PSR) numbers correspond to the following pulsars 
B1054$-$62, J1809$-$1917, B1822$-$14, B1823$-$13, B1828$-$11 consequently.}
\begin{tabular}{ccrrccc} \hline PSR & a & \multicolumn{1}{c}{b} &
\multicolumn{1}{c}{c} & $\chi^2$ & $\nu_{p}^{K}$ & $\nu_{p}$ \\
    &   &  & & &   &  \\
\hline
1 & -7.28$\pm 1.98$ & -1.95$\pm 0.50$  & 1.94$\pm 0.07$  & 4.40 & 1.0 & 0.73 \\
2 & -3.96$\pm 1.14$ & 2.80$\pm 0.84$ & -0.10$\pm 0.09$ & 2.91 & 1.7 & 2.25 \\
3 &  -1.30$\pm 0.29$ & 0.40$\pm 0.20$ & 0.41$\pm 0.03$  & 1.59 & 1.4 & 1.43\\
4 &  -1.18$\pm 0.22$ & 0.52$\pm 0.17$ & 0.56$\pm 0.04$ & 2.35 & 1.6 & 1.66 \\
5 &  -2.09$\pm 0.24$ & -0.55$\pm 0.11$ & 0.12$\pm 0.03$ & 1.69 & 1.2 & 0.74 \\
\hline
\end{tabular}
\end{table}

\section{Be star/pulsar binary and PSR B1259$-$63 spectrum evolution}
\begin{figure}
\includegraphics[width=8cm]{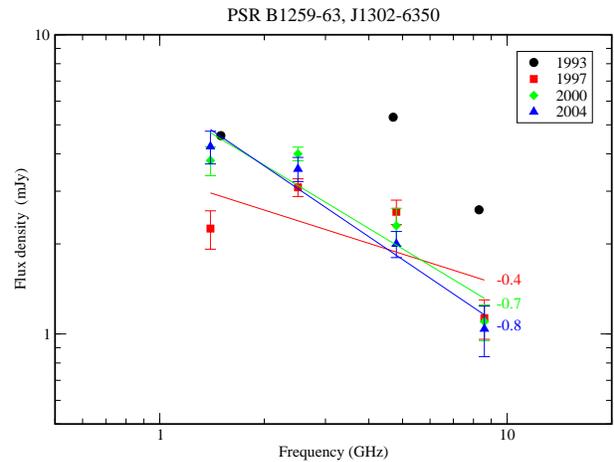}
\caption{The average spectra of PSR B1259$-$63 obtained for three different periastron passages
(1997, 2000 and 2004). Three black dots represent the flux measurements obtained in 1993 by \citet{ma95}.}
\label{fig2}
\end{figure}

To demonstrate the unusual spectral behaviour of B1259$-$63 let us fit its spectra applying the typical power law
method. Using the available measurements of the pulsed flux \citep{ma95, joh99, con02, joh05}, we have constructed
spectra of PSR B1259$-$63 for few different orbital epochs (see Fig. \ref{fig2}). The obtained values of spectral
indices differ significantly from the typical value of about $-1.8$ \citep{maron00}. However, Fig.~\ref{fig2} shows
that this approximation is apparently inaccurate as the spectra clearly show the peaked frequency for the corresponding
periastron passages (see Fig. \ref{fig1} for comparison). Therefore, we have decided to study variation of the observed
radio flux with orbital phases.

\begin{figure}
\includegraphics[width=73mm,angle=0.]{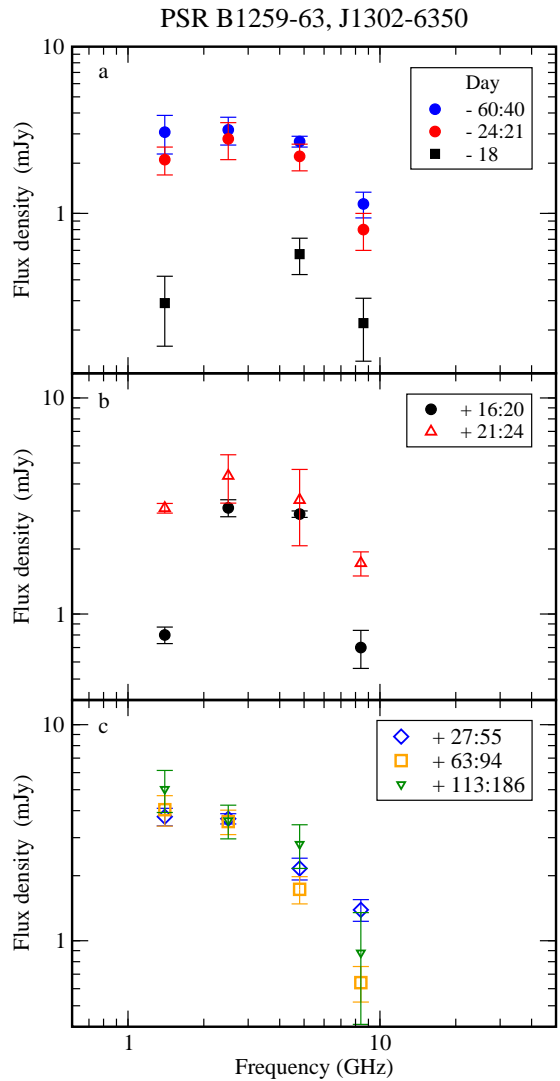}
\caption{The spectra of PSR~B1259$-$63 averaged over each orbital epoch, from 60 days prior to periastron (panel a) up
to 186 days after it (panel c).} 
\label{fig3}
\end{figure}

Using the same database (1997, 2000 and 2004) we have calculated an average flux density for four given frequencies and
for each chosen interval of orbital phases. Fig.~\ref{fig3} shows the spectra for various orbital phase ranges,
determined by the chosen days prior to or past the periastron passage. It is clear that the flux at the given frequency
apparently changes with orbital phases. When the pulsar is close to periastron, the flux generally decreases at all
observed frequencies, but its drastic decrease is observed at the lowest frequency. The spectrum for the furthest
orbital epoch during of which the pulsar was observed so far (i.e. from 113 to 186 days past periastron; see the green
thickened segment in Fig. \ref{fig4}, bottom panel) is consistent with a typical pulsar spectrum described by a
power-law function (see Fig. \ref{fig3} panel c and Fig. \ref{fig4} left panel). On the other hand, the
spectral index for this orbital epoch is of about $-0.8$ which is less than the average value ($-1.8$) for the pulsar
population and one can argue that it rather resembles a broken type spectrum \citep{maron00}. Thus, it cannot be ruled
out that this spectrum is still affected by the stellar wind. Observations of the pulsar near the apastron passage can
clarify this problem.

\begin{table}
\caption{The results of our spectral fits to the PSR~B1259$-$63 flux densities, averaged over the given orbital epochs,
determined by the chosen days prior to or past the periastron passage (see also Figs~3~and~4). There are no errors in
the fit for the orbital phase corresponding to day $-18$ , because in this degenerate case, all errors are zero by
definition (numbers of data points and parameters are the same, see Fig. \ref{fig2}, panel a). }
\begin{tabular}{cccccc} \hline days & a & b & c &  $\chi^2$ & $\nu_{p}$ \\
    &   &  & &   & (GHz) \\
\hline -60 : -40 & -1.70$\pm 0.46$  & 1.38$\pm 0.54$ & 0.27$\pm 0.16$ & 0.74 &
2.54  \\
-24 : -21 & -2.10$\pm 0.22$  & 1.76$\pm 0.23$ & 0.10$\pm 0.05$ & 0.10 &
2.62  \\
-18    & -2.77  & 2.84 & -0.89 & & 3.26 \\
16 : 20   & -4.59$\pm 0.03$  & 4.84$\pm 0.02$ & -0.71$\pm 0.01$~ & 0.01 &
3.37  \\
21 : 24   & -1.72$\pm 0.05$  & 1.51$\pm 0.06$ & 0.31$\pm 0.01$ & 0.01 &
2.75  \\
27 : 55   & -0.88$\pm 0.35$  & 0.34$\pm 0.38$ & 0.55$\pm 0.09$ & 1.49 &
1.57  \\
63 : 94   & -1.48$\pm 0.10$  & 0.55$\pm 0.11$ & 0.56$\pm 0.03$ & 0.04 &
1.53 \\
\hline
\end{tabular}
\end{table}

To analyse the spectrum evolution in more details we have used the method of fitting described in Section 2.
Consequently, we have calculated fitting parameters for function (1) and constructed radio spectra for the chosen
orbital epochs. The fitting parameters,  the reduced $\chi^2$ and the calculated peak frequencies are presented in
Table 2, while the shapes of the spectrum are presented in Fig.~\ref{fig4}, left and right panels. Let us note
that we have used an additional black dashed line in the right panel of Fig.~\ref{fig4} to indicate a fit with an
additional predicted data point which corresponds to the probable flux at frequency 2.6 GHz as there are no
observations for this particular orbital phase (+18 day) at this frequency. The solid black line corresponds to the fit
obtained using only three data points from Fig.~\ref{fig3}. On the other hand one can see that the flux at 2.6 GHz
does not change significantly for all orbital phases (see Fig. \ref{fig3}). Thus, in order to use the same kind of
fitting for all chosen orbital epochs we have calculated the extrapolated flux at 2.6 GHz based on a power law fitting
of two data points at higher frequencies (Fig.~\ref{fig3}, panel a).

The small (much less than unity) values of the reduced $\chi^2$ in Table 2 are caused by rather large
uncertainties in the flux measurements for four orbital epochs (see Fig.3 for comparison). Let us note, that $\chi^2$
is especially small (of about $0.01$) for $+16:20$ and $+21:24$  epochs when the relative uncertainties of the measured
flux are quite large (see Fig. 3b). On the other hand, in the case of epochs when the uncertainties (for all
frequencies) are relatively small (see the blue filled dots in Fig.~3a and the blue empty diamonds in Fig.~3c) the
value of $\chi^2$ is close to unity. We hope to reduce the uncertainties significantly by gathering more flux
measurements during these orbital epochs.

To clarify the spectrum evolution we have presented the corresponding orbital epochs in the bottom panel of
Fig.~\ref{fig4}. Our analysis shows that the shape of the PSR~B1259$-$63 spectrum depends on the orbital phase and
therefore it definitely undergoes evolution. It should be underlined that the peak frequency also depends on the pulsar
orbital phase. Comparing Figs~\ref{fig1} and \ref{fig4} (left and right panels) we can conclude that the spectra of
B1259$-$63 resemble those of the GPS pulsars. Moreover, we can see that the shapes of the B1259$-$63 spectrum at
different orbital phase intervals mimic those of various gigahertz-peaked spectra.
\section{Discussion}
Multi-wavelength observations showed that some pulsars with turn-over in the spectra at high frequencies have very
interesting environments. For example, PSR B1054$-$62 lies behind or within a dense HII region \citep{kori95}, while
PSR B1823$-$13 surroundings appear to show some peculiar properties in radio \citep{gaen03}, as well as in X-ray
observations \citep{pav08}, which may indicate the existence of a compact Pulsar Wind Nebula (PWN). The same holds for
the GPS pulsar J1809$-$1917 \citep{kp07}. This could suggest that the phenomenon of the turn-over at high frequencies
is associated with environmental conditions around the neutron stars rather than with the radio emission mechanism.
Because of more or less stable environmental conditions around the GPS pulsars the shapes of their spectra do not vary
on observable timescales. But in the case of B1259$-$63, the pulsar environment considerably changes due to high
orbital eccentricity while it goes through various orbital phases. The distance of the pulsar from the Be star varies
from 4.8 AU in apastron to 0.34 AU in periastron (by a factor of 14). Be stars have a strong stellar wind and possibly
a strong magnetic field and both the disk density and the magnetic field decrease as the distance from the star
increases. Thus, it is natural to expect that the inhomogeneous environment alters the spectrum of B1259$-$63 in
different ways.

\begin{figure*}
\begin{tabular}{lr}
{\mbox{\includegraphics[width=70mm,angle=0.]{fig4_l.eps}}}&
{\mbox{\includegraphics[width=70mm,angle=0.]{fig4_r.eps}}}\\
\multicolumn{2}{c} {\mbox{\includegraphics[width=87mm,angle=0.]{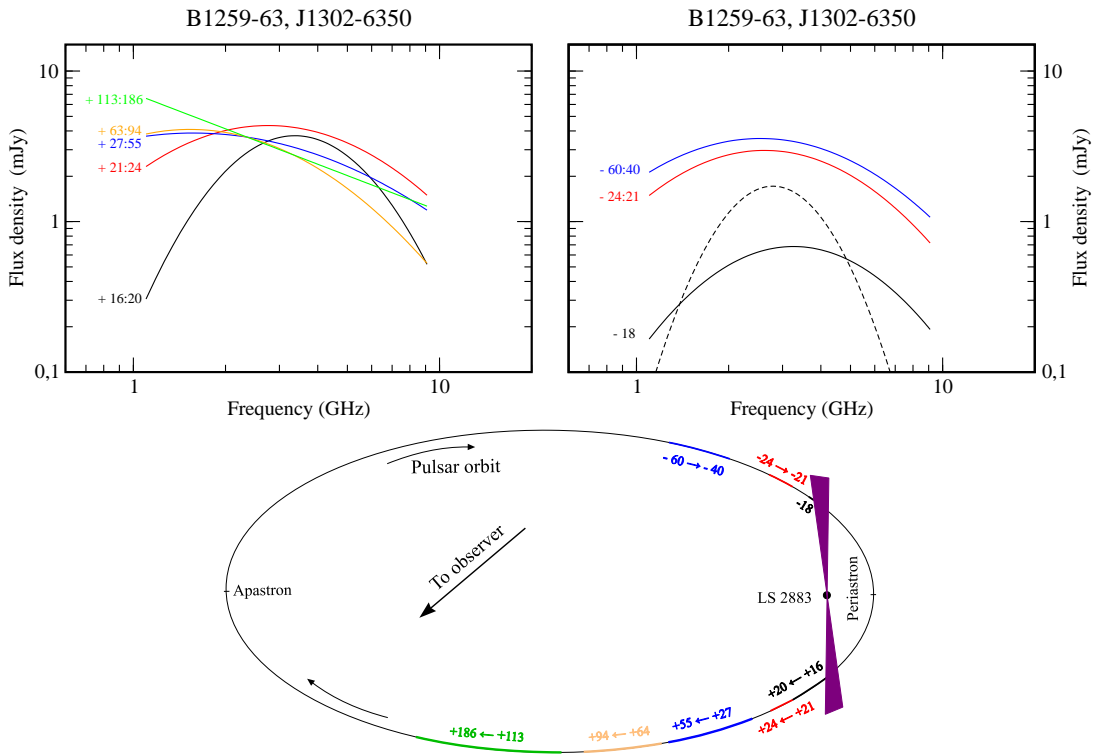}}}
\end{tabular}
\caption{The fits to the B1259$-$63 spectra for the each orbital phase range (left and right panels). 
The corresponding
data points are given in Fig.~\ref{fig3} and the fitted parameters are presented in Table 2. Bottom panel
schematically illustrates the orbital epochs over which the spectra are averaged.} 
\label{fig4}
\end{figure*}

At present it is difficult to construct a detailed theory of the spectra evolution as we lack observational data
especially near apastron (see Fig. \ref{fig4}). These data must be important because the pulsar spectrum is the
least affected by the environment near apastron and this spectrum could be used as a reference spectrum. However, from
the available limited observations we can still deduce general conclusions about the main factors that influence
variations of the observed spectra.

Generally, both the hot stellar wind (by means of thermal absorption) and the magnetic field may provide absorption at
low frequencies \citep{sieb73, km97, kmg00}. While the eclipse itself can be naturally explained by free-free
absorption in the stellar disk \citep{mjm95}, the disk alone is not enough to explain the spectra evolution.

The sign of the GPS feature first appears as early as about 60 days prior to periastron and it is still observed during
at least 90 days after periastron. It is clear that the stellar disk cannot extend that far. It seems that the polar
wind is the only factor that can affect the spectrum of the pulsar. Preliminary estimations show that free-free
absorption due to the stellar wind can be responsible for this effect before the eclipse as well as after it (see Fig.
\ref{fig4}). The optical depth $\tau_{ff}$ of free-free absorption can be expressed as \citep{Ryb79}:
\begin{equation}
\tau_{ff}=0.4\times T_{3}^{-3/2}\nu_{\rm GHz}\int n_{e}^{2}dl_{\rm AU},
\end{equation}
where $n_e$ is the electron density in cm$^{-3}$, $T_3$ is the temperature in $10^3$ K, $\nu_{\rm GHz}$ is the
frequency in GHz and $l_{\rm AU}$ is the distance in AU. The exact value of $\tau_{ff}$ depends on the geometry of the
binary system as well as on the density and temperature of the stellar wind. One can note that the shapes of spectra
are not fully symmetric with respect to the periastron point. To some extent, the spectrum plotted by the green curve
in the left panel of Fig. \ref{fig4} can be used as a reference spectrum. This spectrum is obtained while the
pulsar is in the orbital epoch designated by the thickened green segment in the bottom panel of Fig. \ref{fig4}.
It is the furthest orbital phase at which the pulsar is observed so far, but the spectrum still does not look exactly
like a typical pulsar spectrum. Indeed, the waves emitted before the eclipse travel longer distance through the stellar
wind, thus they are stronger attenuated by free-free absorption, than the waves emitted after the eclipse (see Fig.
\ref{fig4}). This is in agreement with DM and RM variations presented by \citet{mjm95}. The most significant
change of the spectrum shape occurs during a couple of days just before the eclipse as well as immediately afterwards
(the thickened black segments in the bottom panel, Fig.~\ref{fig4}). One can see that the difference between the
red and black curves on the left as well as on the right panel of Fig.~\ref{fig4} is much more noticeable than the
difference between the red and blue curves in the same panels.

In addition, we have shown (see Table 2) that the peak frequency (as it follows from the fitting procedure) also
depends on the orbital phase. This effect suggests that the peak frequency varies with the changes of the pulsar
environment. We argue that such behaviour can be explained by the radio-wave absorption in the magnetic field
associated with the disk. In this case the radio waves should pass through the `magnetosphere' of the disk. Magnetic
field lines just above the stellar disk are populated by the electrons and positrons of the pulsar wind. Such kind of
environment affects the pulsar spectra in the same way as it does in the case of those eclipsing pulsars whose eclipse
duration depends on the observed frequency \citep{km97, kmg00}. The frequency of waves which are affected by the
cyclotron dumping can be expressed in the following way:
\begin{equation}
\nu_{\rm GHz}\approx 2.8\times 10^{-3} \frac{B_0}{\gamma_p (1-\cos\theta)},
\end{equation}
where $B_0$ is the value of the magnetic field associated with the disk, $\gamma_p$ is the Lorentz factor of the
secondary electron-positron pairs of the pulsar wind and $\theta$ is the angle between the wave vector and magnetic
field direction.

Finally let us underline that the PSR B1259$-$63/LS~2883 binary system can hold the clue to the understanding of
the GPS pulsars. As we mentioned in Section 3 the spectrum of B1259$-$63 at the various orbital phases mimics that of
the pulsars with GPS. Thus, we can conclude that the GPS feature should be caused by some external factors rather than
by the emission mechanism. On the other hand, the GPS pulsars are isolated radio pulsars and therefore, we cannot draw
a direct analogy between the PSR B1259$-$63/LS~2883 system and the the GPS pulsars, as the latter have no companion
stars and/or disks. But the GPS pulsars apparently are surrounded by some kind of environment that can affect the
spectra of those pulsars in the same way as the stellar wind affects the B1259$-$63 spectrum. As an example we can take
the cases of B1823$-$13 which is surrounded by a compact pulsar wind nebula \cite{pav11}, or PSR B1054-62 which lies
behind or within a dense HII region \citep{kori95}. All GPS pulsars have relatively high Dispersion Measures that in
some cases, are too large to be accounted for by the Galactic electron density, and thus, we can speculate that there
is a quite high particle density in the vicinity of these pulsars \citep[see also][]{kijak11}. As soon as all the
necessary observational data are available, we will model the physical conditions which cause the evolution of the
B1259$-$63 spectrum. Consequently, we will be able to estimate the particle density, temperature and magnetic field
that are necessary to form the spectra shown in Fig.~\ref{fig4}. We believe that the same physical processes (i.e.
free-free and/or cyclotron absorption) are responsible for both B1259$-$63 and the GPS pulsars spectra, therefore, we
will be able to estimate characteristic values for the GPS pulsar surroundings in a similar way. The only difference
could be an invariable shape of the gigahertz-peaked spectra, in contrast with the B1259$-$63 spectrum, which undergoes
evolution due to orbital motion.

\section{Summary}

Finally, we can conclude that the B1259$-$63 spectrum undergoes evolution while the pulsar orbits around the Be star LS
2883. The most significant change of the spectrum shape occurs during a couple of days just before the eclipse as well
as immediately afterwards (the thickened black segments in the bottom panel, Fig.~\ref{fig4}). We argue that the
observed variation of the spectra is caused by a combination of two effects: the free-free absorption in the stellar
wind and the cyclotron resonance in the magnetic field. This field is associated with the disk and is infused by the
relativistic particles of the pulsar wind. Having noticed the apparent resemblance between the B1259$-$63 spectrum and
the GPS, we suggest that the same mechanisms should be responsible for both cases. Thus, the case of B1259$-$63 can be
treated as a key factor to explain the GPS phenomenon observed for the solitary pulsars with interesting environments.
Therefore, the binary system B1259$-$63/LS 2883 seems to be an important astrophysical laboratory to study the
interaction between pulsar and their environments, such as bow-shocks and pulsar wind nebulae \citep{pav11}.

\section*{Acknowledgments}
We are grateful to anonymous referee for useful comments. This work was partially supported by the the Polish State
Committee for Scientific Research under Grant N N203 391934. GM was partially supported by the Georgian NSF grant ST08/4-442.                                                     
We thank M.Margishvili for the language editing of the manuscript.


\begin{thebibliography}{99}
\bibitem[\protect\citeauthoryear{Connors et al.}{2002}]{con02} Connors T. W., Johnston S.,
Manchester R.~N., McConnell D., 2002, MNRAS, 336, 1201
\bibitem[\protect\citeauthoryear{Gerhz \& Woolf}{1970}]{b7} Gerhz R.D., Woolf N.J.,
1970, ApJ, 161, L213
\bibitem[\protect\citeauthoryear{Gaensler et al.}{2003}]{gaen03} Gaensler B. M.,
Schulz N. S., Kaspi V. M., Pivovaroff M. J., Becker W. E., 2003, ApJ, 588, 441
\bibitem[\protect\citeauthoryear{Hobbs et al.}{2004}]{hobs04} Hobbs G. et al., 2004, MNRAS, 352, 1439
\bibitem[\protect\citeauthoryear{Johnston et al.}{1992}]{joh92} Johnston S., Lyne A. G., Manchester R. N.,
Kniffen D. A., D'Amico N., Lim J., Ashworth M., 1992, MNRAS, 255, 401
\bibitem[\protect\citeauthoryear{Johnston et al.}{1994}]{joh94}Johnston S., Manchester R. N., Lyne A. G.,
Nicastro L., Spyromilio J., 1994, MNRAS, 268, 430
\bibitem[\protect\citeauthoryear{Johnston et al.}{1999}]{joh99} Johnston S., Manchester R. N., McConnell D.,
Campbell-Wilson D., 1999, MNRAS, 302, 277
\bibitem[\protect\citeauthoryear{Johnston et al.}{2005}]{joh05}Johnston S., Ball L., Wang N.,
Manchester R. N., 2005, MNRAS, 358, 1069
\bibitem[\protect\citeauthoryear{Kargaltsev \& Pavlov}{2007}]{kp07} Kargaltsev O., Pavlov G. G., 2007, ApJ, 670, 655
\bibitem[\protect\citeauthoryear{Khechinashvili \& Melikidze}{1997}]{km97} Khechinashvili D., Melikidze G., 1997, A\&A, 320, L45
\bibitem[\protect\citeauthoryear{Khechinashvili et al.}{2000}]{kmg00} Khechinashvili D., Melikidze G., Gil J., 2000, ApJ, 541, 335
\bibitem[\protect\citeauthoryear{Kijak \& Maron}{2004}]{km04} Kijak J., Maron, O., 2004,
in Camilo F., Gaensler B.M., eds, Proc. IAU Symp. 218, Young Neutron Stars and Their Environments.
Astron. Soc. Pac., San Francisco, p. 339
\bibitem[\protect\citeauthoryear{Kijak et al.}{2007}]{kijak07} Kijak J., Gupta Y., Krzeszowski K., 2007, A\&A, 462, 699
\bibitem[\protect\citeauthoryear{Kijak et al.}{2011}]{kijak11} Kijak J., Lewandowski W., Maron O., Gupta Y., Jessner A., 2011, A\&A, 531, A16
\bibitem[\protect\citeauthoryear{Kong et al.}{2011}]{kong11} Kong S. W., Yu Y. W., Huang Y. F., Cheng K. S.,
2011, MNRAS, in press
\bibitem[\protect\citeauthoryear{Koribalski et al.}{1995}]{kori95} Koribalski B.,
Johnston S., Weisberg J. M., Wilson W., 1995, ApJ, 441, 756
\bibitem[\protect\citeauthoryear{Kuzmin \& Losovsky}{2001}]{kuzm01} Kuzmin A. D., Losovsky B. Ya., 2001, A\&A, 368, 230
\bibitem[\protect\citeauthoryear{Lorimer et al.}{1995}]{lori95} Lorimer D. R.,
Yates J. A., Lyne A. G., Gould D. M., 1995, MNRAS, 273, 411
\bibitem[\protect\citeauthoryear{Malofeev et al.}{1994}]{mal94} Malofeev V. M., Gil J., Jessner A.,
Malov I. F., Seiradakis J. H., Sieber W., Wielebinski R., 1994, A\&A, 285, 201
\bibitem[\protect\citeauthoryear{Manchester \& Johnston}{1995}]{ma95} Manchester R. N., Johnston S., 1995, ApJ, 441, L65
\bibitem[\protect\citeauthoryear{Maron et al.}{2000}]{maron00} Maron O.,
Kijak J., Kramer M., Wielebinski R., 2000, A\&A S, 147, 195
\bibitem[\protect\citeauthoryear{Melatos et al.}{1995}]{mjm95} Melatos A., Johnston S., Melrose, D., 1995, MNRAS, 275, 381
\bibitem[\protect\citeauthoryear{McLaughlin et al.}{2002}]{mcla02} McLaughlin M. A. et al.,
2002, ApJ, 564, 333
\bibitem[\protect\citeauthoryear{Pavlov et al.}{2008}]{pav08} Pavlov G.G., Kargaltsev O., Brisken W.F., 2008, ApJ, 675, 683
\bibitem[\protect\citeauthoryear{Pavlov et al.}{2011}]{pav11} Pavlov G. G., Chang, C., Kargaltsev O., 2011, ApJ, 730, 2p
\bibitem[\protect\citeauthoryear{Rybicki \& Lightman}{1979}]{Ryb79} Rybicki G. P., Lightman A. P., 1979, "Radiative processes in Astrophysics", Wiley-Interscience, New York
\bibitem[\protect\citeauthoryear{Sieber}{1973}]{sieb73} Sieber W., 1973, A\&A S, 28, 237
\end{thebibliography}
\end{document}